# An Australian DER Bill of Rights and Responsibilities

N.N. Lal[1*], L. Brown[2]

1. College of Engineering, Australian National University, Acton, ACT 2600

2. 5 North Hidden Valley Cct, Beaconsfield, VIC 3807, Australia

https://doi.org/10.48550/arXiv.2112.04855  | Cite as arXiv:2112.04855

*corresponding author: niraj.lal@anu.edu.au

**Abstract**

Australia's world-leading penetration of distributed solar photovoltaics (PV) is now impacting power system security and as a result how customers can use and export their own PV-generated energy. Several programs of Australian regulatory reform for distributed energy resources (DER) have emphasised the importance of placing consumers at the centre of any energy transition, but this has occurred against a haphazard backdrop of proposals for solar export taxes, updated inverter standards, and diminishing feed-in-tariffs. Absent from the discussion is a coherent espousal of reasonable consumer expectations with practical technical definitions of how these may be applied. Whilst American legislation has enshrined initial rights to connect PV, they do not consider the evolution of rights in a DER-*dominated* future. Operating a power system with high (>70%) penetration of DER will likely require increased control of customer owned assets with additional requirements for forecasting to maintain system security, reliability and equity. This paper proposes a first attempt at a 'DER Bill of Rights and Responsibilities' for both passive and active participation in energy markets, using Australia as an example jurisdiction. Guiding principles of fairness are presented with practical definitions referencing existing instruments including inverter standards, network connection agreements, the reliability standard, and central ancillary service markets. The intent is that clarity on rights to self-consumption and passive participation will support customer trust in the guiderails of active DER integration and control. We highlight how these proposed rights are already being breached regularly in Australia, before outlining a pathway to enshrine them for a DER-dominated future with broad sector endorsement. These questions are critical for Australia to address now; it is likely other countries will be required to do so in the near future.

## 1. Introduction

### 1.1 Context – DER penetration is now impacting grid security in Australia

The world solar revolution is underway. Solar panels are now cheaper per square metre than marine-grade plywood (ITRPV 2021). Swanson's law continues to show exponential decline in price with cumulative global capacity. More solar photovoltaics (PV) is being installed globally than all other forms of generation (IRENA 2021), likely to accelerate for the coming three decades towards 2050 as humanity decarbonises its energy sector across the planet.

Australia – whilst being amongst the world's largest exporters of both coal and gas, surprisingly finds itself with by far the world's most installations of solar PV per capita (Stocks 2021). Beyond providing bragging rights, the per-capita highest ranking of solar PV penetration is now critically impacting system security, displacing thermal generators that provide minimum secure levels of inertia and system-strength, regularly breaking minimum system load records (AEMO 2021), and breaching normal operating voltages on the distribution network (Stringer 2021).

To maintain system security and reliability with very high (70%) DER penetration, system operators will almost certainly require increased control and curtailment of customer-owned assets (particularly of exported energy, though at times total PV output) and increased requirements for visibility and forecasting (particularly for algorithm-controlled charging and discharging of energy storage). To address this, decisions are being made now about emergency backstop PV control measures, solar export charges, dynamic operating envelopes, flexible trader arrangements, and myriad other regulatory reforms for the use, control, connection and engagement of solar PV and the broader range of distributed energy resources (DER) including smart devices, home batteries and electrical vehicles. The social contract of connecting to the grid is implicitly evolving beyond passive expectations of electricity provided under reliability standards established in the 19$^{th}$ century and minimal customer responsibilities of purchasing products that conform to technical standards established in the 20$^{th}$ century, to a two-way engagement to support fairness of supply for both DER-owners and non-owners in a grid that is at times dominated by DER.

Whilst there are updates to Access and Pricing Arrangements for DER (AEMC 2021), application of DER market and Virtual Power Plant (VPP) trials, reviews of the National Energy Customer Framework and development various industry codes of conduct (Clean Energy Council 2021), there is absent a clear, intelligible communication of customer rights and responsibilities for participation in the future energy system. Particularly absent is a delineation between active engagement with energy markets e.g., through remunerated exported of solar PV electricity, and passive engagement with electricity such as the use of a dialysis machine.

## 1.2 Article outline

This article presents a draft DER Bill of Rights and Responsibilities to support an environment of trust and confidence in the safeguards and guidelines of participation and remuneration for DER integration into the energy system. That is, with a Bill of Rights and Responsibilities which reform initiatives commit to abiding by, customers and their representatives can have confidence that core rights will not be breached, with clear knowledge of technological and participatory responsibilities, all without requiring detailed knowledge about dynamic operating envelopes, essential system services, flexible trader arrangements, parent-child metering definitions, communication protocols, and the wide range of other complex technical issues surrounding DER integration. High-level principles are presented to form the customer-interaction guiderails by which both any product technical design feature or regulatory reform proposal may be assessed. These principles include:

- To support system security and reliability with high DER penetration.
- To preserve the precedent of fair access to energy use by reasonable passive loads.
- To allow the self-use of self-generated electricity with minimum interference or obligation (with curtailment of self-consumption not exceeding the value of unserved energy for passive loads).
- To outline that where customers wish to interact with energy markets for remuneration though import, export or the provision of system services, they may be subjected to additional obligations of technical standards for grid support, information provision, and remote control or disconnection.
- That resources be treated symmetrically with large-scale resources where possible and efficient, and that some resources may be required to connect actively based on threshold kW or kWh values.
- That customers have a right to privacy and access to fair share of value from their energy data.
- That passive options should be set as the default for DER, with informed customer choice required for DER to participate actively in energy markets, and that active options should only be enabled where there is net benefit to the customer.

In this paper, we first outline existing rights and the context of PV and DER integration with a review of recent regulatory reforms. We discuss what these reforms do not yet achieve for customers before outlining the current exposition of international rights for DER which to date has focused on solar PV connection. We then present a draft Australian DER Bill of Rights and Responsibilities accompanied by guiding principles and practical example definitions referencing existing instruments including inverter standards, network connection agreements and central ancillary service markets. Rights and Responsibilities are structured by energy activity (Consumption, Generation/Export, Storage, Data) and interaction type (Passive vs Active). Principles for active interaction may particularly lend themselves to the formulation of future Dynamic Network Agreements. We highlight how proposed rights and responsibilities are already being breached regularly in Australia, before outlining a pathway to enshrine them for a DER-dominated future.

## 2. Review

### 2.1 Existing regulatory frameworks were not designed for high DER penetration

Historically, customers in Australia have been allowed to install 5kW PV systems per phase at their household with an unrestricted ability to export electricity to the grid. But with rapidly rising PV penetration, to manage distribution voltages within safe limits distribution companies are increasingly restricting export limits to 2kW, 1kW or refusing export entirely (AER 2021). In the Australian state of Victoria as an example export levels are being reduced for between 20-40% of new connections for AusNet Services (with approx. 1.5 million customers) and around 20% of connection offers for Powercor and CitiPower (with approx. 1.2 million customers).

In addition to fixed export limits being applied at time of connection, solar PV systems are increasingly being automatically curtailed due to standard grid support settings on PV inverters (AS 4777.2) in response to high distribution network voltages (Stringer et al. 2021), with a loss in value to customers of $1.2-4.5m per year in South Australia alone. These responses not only reduce the amount of exported electricity, but *the entire output* of solar PV inverters. That is, when network voltages are high, customers are not able to self-consume their own generated electricity.

Further to distribution-network challenges the rapid increase of solar PV is causing minimum system load to fall to record levels approaching minimum secure limits (AEMO 2021). This has led the Australian Energy Market Operator to implement Minimum System Load procedures with actions ranging from 1) advance notice of minimum system loads, 2) operational actions to maintain system security with minimum system loads, and 3) curtailment of rooftop solar PV (AEMO 2021b). Whilst the aim is that curtailment is used only rarely by AEMO there are no guidelines for ceiling levels of use (in contrast to clearly defined reliability standards), nor is it clear how the switch-off capability is intended to be used by future distribution system operators. The solar curtailment measures also prevent customers from being able to self-consume their own generated electricity.

The main protections for Australian customer engagement with energy are found in the federal government's New Energy Customer Framework (NECF) (AEMC 2021b) which "regulates the connection, supply and sale of energy (electricity and gas) to grid-connected residents". This framework is comprised of the National Energy Retail Law, the National Energy Retail Regulations and the National Energy Retail Rules, alongside the Australian Consumer Law. It is arguable whether this framework provides clear, intelligible or adequate protection for customers in the access and purchasing of retail electricity, but the framework was certainly not designed for customers with solar PV, batteries or other active DER in mind.

In addition to the Framework there exists the New Energy Tech Consumer Code written predominantly by solar PV retailers with the Clean Energy Council and Smart Energy Council to provide an industry code to self-regulate sales conduct (Clean Energy Council 2021). The Australian Energy Charter to "progress the culture and solutions needed" by CEOs of Australian energy businesses to meet customer expectations for a sustainable energy future (Energy Charter 2019). There also exists the Australian Consumer Data Right for Energy which aims to support customer access, privacy and control of energy data and support easier retail switching (ACCC 2021). Further work that references customer rights with DER includes the Energy Security Board's DER Maturity Plan (ESB 2021), ARENA's Distributed Energy Integration Program (ARENA 2021), and AEMO's Engineering Framework (AEMO 2021c).

## 2.2 Vital work is progressing on fair, equitable solar PV export arrangements

In 2021 the Australian Energy Market Commission made its final determination on Access and Pricing and Incentive Arrangements for Distributed Energy Resources (AEMC2021b) outlining i) Obligations on DNSPs to provide export services (distributors must offer a 'basic' export level, and avoid static zero export limits where possible), ii) Enabling new network tariff options that reward customers (DNSP are now allowed to charge for export), and iii) Strengthening consumer protections and regulatory oversight (the Australian Energy Regulator is now required to regularly review the provision of export services and reasonable Customer Export Curtailment Values).

These are important steps to reduce the inequitable impacts of solar PV integration on customers that don't (or can't) own solar panels. Proponents of the rule change were the Australian Council of Social Services, the Total Environment Centre, St Vincent de Paul Society Victoria and SA Power Networks, and they presented a compelling case for fair distribution of benefits from DER, and in particular customer rights to export (AEMC 2021b).

**But there is no clear framework of customer rights and obligations for a DER-*dominated* future**

Whilst the Access Pricing and Incentive Arrangements provide some clarity on direction, as the penetration of solar PV increases from the current 1 in 4 households to beyond, and as battery and electric vehicle adoption accelerates alongside broader household electrification and increasing market-connectivity, there is a need for a clearer, more intelligible, shared understanding of fair customer-system interaction in a DER-dominated future.

Actively controllable DER load and generation will likely comprise at times the majority of load and generation on the future system. In this future, a basic non-zero export level may be difficult or inefficient to facilitate without some form of a dynamic operating envelope. Distributed solar PV is already meeting >60% of generation in South Australia at times (AEMO 2021). With the introduction of several market reforms such as the Wholesale Demand Response Mechanism and Scheduled Lite to provide additional visibility of non-scheduled load, it is likely the system will also evolve over time towards significant penetration of controllable load (ESB 2021). How should this future active DER interaction be regulated? How will the future system fairly accommodate this active interaction with the right to passive interaction? How might this be structured so as to fairly distribute the benefits of DER penetration and not unjustly penalise customers that do not have access to DER?

From a customer's perspective, what should a customer's right be to self-consume the electricity they've generated? What are the implicit obligations for all consumers arising from connection to the public asset of the electricity grid? What rights and obligations should accompany customers who use active DER loads or generation that respond to market signals? What are reasonable customer expectations of solar PV connection and basic export? What are the minimum technical capabilities that products should have to allow passive/active DER connection? What are responsible guidelines regarding interaction with Dynamic Operating Envelopes and the setting of reasonable Customer Export Curtailment Values?

Current regulatory frameworks do not clearly address these questions nor provide an intelligible set of customer rights and obligations by which any reform might be measured against. This article aims to redress this balance. By doing so it aims to support the communication of customer rights and responsibilities for participation in the energy system with active (that is market-responsive) distributed energy resources.

There is a distinction here between active DER and passive DER (and customer loads) that do not engage with the energy market nor aim to be controllable or discretionary. Any new framework should respect expectations of reasonable passive engagement with electricity – to be able to plug normal things into the wall and expect them to work and not pay too much for them. What is normal? This is further outlined below but perhaps might best be summed up as load and/or generation resources that aren't remotely controllable and that don't respond to market signals.

This principle of distinguishing between active and passive DER is potentially supported by Flexible Trader Arrangements currently being considered by Australian market bodies, where a separate 'child' meter (either physical or virtual) coordinates active DER behind the 'parent' meter – allowing an aggregator or Trader to control and settle two-way energy flows and services on behalf of a customer.

The aim for a Bill of Rights and Responsibilities is to support sector reform for the broad range of technical issues regarding DER integration to occur in an environment of trust and confidence in safeguards and guidelines of participation and remuneration. With clearly outlined principles for energy consumption, export, technical capability, and market participation, customers and their representatives can have confidence that core rights will not be breached, with clear knowledge of responsibilities, without requiring detailed technical knowledge about dynamic operating envelopes, essential system services, flexible trader arrangements, parent-child metering definitions, data access models, or communication protocols.

**2.3 International progress is helpful for rights of PV ownership and connection in a DER-minority grid**

Jurisdictions in the USA have progressed furthest to enshrine rights and responsibilities of solar PV and DER more broadly, though progress has not always been smooth. In 2019, two California Senators introduced a Solar Bill of Rights bill SB-288 to prevent monthly fees from being charged to solar PV owners, amongst other reforms, but the final bill that was passed did not include this proposal and further removed all references to 'Solar Rights' (State of California 2019). In August 2021 New York introduced legislation A1933 to prohibit Homeowner Associations from adopting "any rules or regulations that would effectively prohibit, or impose unreasonable limitations on, the installation or use of solar power systems". Unreasonable limitations include (1) Limitations that inhibit the solar power system from functioning at maximum efficiency, and (2) Limitations that increase costs greater than 10% of the total cost of the installation of the solar power system (State of New York 2021).

The USA state of Nevada has the most wide-ranging protection of distributed solar PV connection and use in their Renewable Energy Bill of Rights which was passed in 2017 (State of Nevada 2017). Rights include those to:

1. Generate, consume and export renewable energy and reduce his or her use of electricity that is obtained from the grid.

2. Use technology to store energy at his or her residence.

3. If the person generates or stores energy be allowed to connect his or her system:
   (a) In a timely manner;
   (b) In accordance with requirements established by the electric utility to ensure the safety of utility workers; and
   (c) After providing written notice to the electric utility providing service in the service territory and installing a nomenclature plate on the electrical meter panel indicating that a system that generates renewable energy or stores energy, or any combination thereof, is present if the system:
     (1) Is not used for exporting renewable energy past the electric utility meter on the customer's side; and
     (2) Meets all applicable state and local safety and electrical code requirements.

4. Fair credit for any energy exported to the grid.

5. Consumer protections in contracts for renewable energy.

6. Have his or her generation of renewable energy given priority in planning and acquisition of energy resources by an electric utility.

7. Remain within the existing broad rate class to which the resident would belong in the absence of a net metering system or a system that generates renewable energy or stores energy, or any combination thereof, without any fees or charges that are different than the fees and charges assessed to customers of the same rate class, regardless of the technologies on the customer's side of the electricity meter, including, without limitation, energy production, energy savings, energy consumption, energy storage or energy shifting technologies, provided that such technologies do not compromise the safety and reliability of the utility grid.

This legislation has been a leading step in protecting the ownership, connection and use of PV electricity in Nevada and elsewhere and comprises a fit-for-purpose framework for distributed PV interaction in grids with relatively low penetration. There are approximately 70,000 solar PV systems in Nevada, totalling 400 MW capacity, in a grid which supplies 3.1million customers with a peak demand of approximately 12GW.

In contrast, the state of South Australia alone has more than 300,000 solar PV systems, totalling more than 1.5 GW capacity, in a grid supplying 1.7 million customers with a peak demand less than 3.5 GW. Distributed solar PV has recently provided 88% of total demand in South Australia, with total solar PV (including utility-scale) regularly meeting 100% of the state's demand (AEMO 2021). The rest of the states in Australia are rapidly catching up.

**4. An Australian DER Bill of Rights and Responsibilities for a DER-dominated grid**

Whilst helpful in articulating rights of customers to connect and use solar PV systems, the existing legislative frameworks, including in the USA, do not consider the evolution of customer rights in a DER-*dominated* future where DER can comprise a majority of grid-scale generation.

In this rapidly emerging paradigm, there are two qualitative differences for DER: 1) export curtailment will be necessary at times to ensure i) network and system security, and ii) just distribution of benefits for other customers; and 2) DER will be required to provide system services at a minimum via passive connection settings, and increasingly via active interaction where efficient to do so. The implicit obligations on a customer connecting to the public asset of the electricity grid must now broaden to encompass responsible DER interaction.

This paper proposes a first attempt at an Australian 'DER Bill of Rights and Responsibilities' in this DER dominated future, for four areas: 1) Energy consumption 2) Energy generation and 3) Energy storage and 4) Energy data.

Each area has rights and responsibilities outlined for two broad categories of interaction 'Passive' and 'Active', and are informed by guiding principles including:

- To support system security and reliability with high DER penetration.
- To preserve the precedent of fair access to energy use by reasonable passive loads.
- To allow the self-use of self-generated electricity with minimum interference or obligation (with curtailment of self-consumption not exceeding the value of unserved energy for passive loads).
- To outline that where customers wish to interact with energy markets for remuneration though import, export or the provision of system services, they may be subjected to additional obligations of technical standards for grid support, information provision, and remote control or disconnection.
- That resources be treated symmetrically with large-scale resources where possible and efficient, and that some resources may be required to connect actively based on threshold kW or kWh values.
- That customers have a right to privacy and access to fair share of value from their energy data.
- That passive options should be set as the default for DER, with informed customer choice required for DER to participate actively in energy markets, and that active options should only be enabled where there is net benefit to the customer.

The bill is summarised in a table and we then discuss practical technical definitions of how these may be applied with references to existing instruments where possible, including inverter standards, network connection agreements and central ancillary service markets. Following this, we highlight how these proposed rights are already being breached regularly in Australia, before outlining a regulatory reform pathway to enshrine them for a DER-dominated future.

We conclude by outlining how simple espousal of these rights would support customers to trust ongoing regulatory reform, and also support the industry by providing clarity on the guiderails of interaction and system design.

**4.1 An Australian DER Bill of Rights and Responsibilities – Summary Table**

Energy customers have the following rights and responsibilities[1] (The full draft Bill is included in Appendix A)

**Table 1. Summary of the draft DER Bill of Rights and Responsibilities**

| Energy activity / Grid interaction | Passive interaction<br>*Energy use without active market participation. Generation and storage for self-consumption.* | Active interaction<br>*Energy use with active market participation. Generation and storage for remunerated export and service provision.* |
|---|---|---|
| **1. Energy Consumption** | Right to connect and consume energy at fair prices for reasonable passive loads that do not participate in energy markets.<br><br>Some loads may be required to be actively connected based on threshold kW/kWh consumption*.<br><br>Electricity supply for passive loads should meet the Reliability Standard. | Right to own active loads that participate in energy markets.<br><br>When connected, active loads may be subjected to obligations to:<br>　i) meet additional technical standards for active grid support<br>　ii) register resource details at time of connection<br>　iii) allow remote control and/or disconnection for |

---

[1]*Rights are enumerated as follows:*

*R[Energy activity].[Grid interaction type].[Right/Responsibility]*

*For example: R2.2.1 indicates [Energy Generation].[Active Grid Interaction].[Right/Responsibility #1]*

| Energy activity / Grid interaction | **Passive interaction** *Energy use without active market participation. Generation and storage for self-consumption.* | **Active interaction** *Energy use with active market participation. Generation and storage for remunerated export and service provision.* |
|---|---|---|
| | Example passive loads include hair-dryers, toasters or dialysis machines.<br><br>*For example, electric vehicle charging points. | credible risks to system security<br>iv) provide good-faith scheduling information to market and/or network operators.<br><br>Right to switch active loads back to passive settings (noting that some loads may be required to be active)<br><br>Active loads are treated and remunerated symmetrically with large-scale loads where possible and efficient.<br><br>Example active loads include electric hot-water storage systems, pool pumps, or electric vehicle charging points that participate in grid demand management programs or are otherwise market responsive. |
| **2. Energy Generation** | Right to install renewable energy generation resources, subject to relevant requirements for safety and quality.<br><br>Right to consume energy generated on-site with minimal restriction.<br><br>Energy generation systems may be subjected to obligations to:<br>    i) meet minimum technical standards for passive grid support<br>    ii) register system details at time of connection<br>    iii) provide visibility of system operation to market and/or network operators<br>    iv) allow remote control and/or disconnection for non-credible risks to system security<br><br>Curtailment of self-consumption to not exceed the unserved energy provision in the Reliability Standard. | Right to own generation systems that participate in energy markets<br><br>Energy generation systems capable of participating in energy markets may be subjected to additional obligations to:<br>    i) meet additional technical standards for active grid support<br>    ii) provide details of system operation<br>    iii) provide forecasts of system operation to market and/or network operators<br>    iv) allow adjustment and curtailment of exported energy to support system and network performance<br>    v) allow remote control and/or disconnection for credible risks to system security<br><br>Generating systems that participate in energy markets have a right to symmetric treatment and remuneration with large-scale |

| Energy activity / Grid interaction | Passive interaction  *Energy use without active market participation. Generation and storage for self-consumption.* | Active interaction  *Energy use with active market participation. Generation and storage for remunerated export and service provision.* |
|---|---|---|
| | An example energy generation system is a solar PV system. | generation where possible and efficient  Right to switch actively participating systems back to passive settings, with minimal restriction of operation. |
| **3. Energy Storage** | Right to install energy storage resources on their premises, subject to relevant requirements for safety and quality.  Right to store energy generated on-site for later self-consumption.  Storage systems may be subjected to obligations to:      i) meet minimum technical standards for grid support      ii) register system details at time of connection      iii) provide visibility of system operation to market and/or network operators  An example energy storage system is a home battery. | Right to own storage systems that participate in energy markets  Where capable of storing energy from the grid, operating as active load or participating in energy markets, storage systems may be subjected to additional obligations to:      i) meet additional technical standards for active grid support      ii) provide visibility of system operation to market and/or network operators      iii) allow remote control and/or disconnection for credible risks to system security      iv) provide good-faith scheduling information to market and/or network operators  Active storage systems include home batteries or electric vehicle batteries capable of charging/discharging from/to the grid. |
| **4. Energy Data** | Right to privacy through their energy data  Right to access their energy data  Right to be informed about the uses of their energy data  Right to a fair share in the value of third-party licensing of their energy data  Right to revoke consent of the use of their energy data | Where customers are actively interacting with the grid, they may have obligations to provide energy data relating to resource and/or system details, current operation, and forecast operation. |

| Energy activity / Grid interaction | Passive interaction<br>*Energy use without active market participation. Generation and storage for self-consumption.* | Active interaction<br>*Energy use with active market participation. Generation and storage for remunerated export and service provision.* |
|---|---|---|
| | Energy data includes consumption data, billing data, resource settings and operational data, and forecasted operational data. | |
| **5. Default Settings** | Passive options are set as default for DER<br><br>Informed customer choice is required for DER to participate actively | Active options should only be enabled where there is net benefit to the customer.<br><br>Right for customers to revert to passive settings<br><br>Passive settings may be reverted to in the event of:<br>  i) Loss of communication with DER<br>  ii) Failure of compliance by DER<br>  iii) Cybersecurity compromise |

## 5. Discussion

### 5.1 Principles

This first draft of a DER Bill of Rights and Responsibilities is constructed with guiding principles of fairness. Firstly, the implicit principle to support system security and reliability with high DER penetration for the benefit of all electricity customers. In Australia this principle is enshrined in the National Electricity Objective a stated in the National Electricity Law (National Electricity [South Australia] Act 1996).

Beyond this, the Bill recognises a principle to preserve the precedent of fair access to energy use by reasonable passive loads. Electricity is an essential service and there is a historical expectation of an ability to connect to the grid and consume electricity in reasonable ways with safety, at fair prices, and with minimal onerous requirements. This is particularly the case for electrical devices that would likely never participate in energy markets or be remotely controlled, such as hair-dryers, toasters or dialysis machines. The Bill preserves this expectation of reasonable use from passive loads in R1.1.1, with a requirement that the provision of electricity should meet the Reliability Standard – in Australia this states that at least 99.998% of forecast customer demand to be met each year (to keep unserved energy to less than 0.002% in any year). There is an even tighter interim measure set by the Energy Security Board for unserved energy to be less than 0.0006% of any year. This requirement of the Reliability Standard is enshrined in R1.1.2.

There is an additional guiding principle of the right to install equipment at one's private property for self-use. We extend this implicit expectation to the self-use of electricity generation on one's own property. Where an individual installs a PV system at their property, subject to relevant requirements for safety and quality, a fair expectation is to be allowed to self-consume any generated electricity with minimum interference and only minimum technical obligations. This expectation should also be applied for storage and the subsequent self-consumption of self-generated electricity. These rights are outlined in R2.2.2 and R3.2.2

Whilst self-consumption of self-generated electricity should be guaranteed where possible, there must be a recognition that when connected to the grid even the self-consumption of onsite generated electricity may have consequences for grid operation (for example contingency frequency control and disturbance ride through). As a result, and especially for a DER-dominated future, there are possible obligations – which are framed as lightly as practicable - on resources to:

>    i) meet minimum technical standards for passive grid support (such as disturbance ride-through settings),
>
>    ii) register system details at time of connection (for example through a DER Register)
>
>    iii) provide visibility of system operation to market and/or network operators (for example kW output of PV systems to support operator visibility and solar forecasting), and
>
>    iv) allow remote control and/or disconnection for non-credible risks to system security (for example in the event of major disturbances).

In symmetry with the supply of energy for passive loads, the curtailment of self-generated electricity may on occasion by necessary for grid operation, but should only occur rarely. We propose to tie this value to the Reliability Standard and the framework of credible- and non-credible risks to system security. In this way, we propose the curtailment of self-generated electricity value not exceed the value for unserved energy for passive loads (R2.1.3).

Responsibilities are similarly outlined in R2.2.3 and R3.1.3 for generation and storage resources respectively. The responsibilities – even for passive interaction – form part of the implicit contract of engagement with the public asset of an electricity grid. By the mere act of connecting to the grid, a customer is benefiting from a public asset and public investment and there are minimum responsibilities that accompany this. We propose that on occasion this also encompasses the right to self-consume self-generated electricity.

The implicit contract of engagement between an energy customer and the grid is proposed by this Bill to change substantially when a customer wishes to participate actively in energy markets. That is, where customers wish to interact with energy markets for remuneration though import, export, or the provision of system services for future renewable-dominated grids (Lal 2021), they may be subjected to additional obligations including to:

>    i) meet additional technical standards for active grid support (for example voltage response modes for PV inverters)
>
>    ii) provide visibility of system operation to market and/or network operators (for example through visibility of kW output or state of charge)
>
>    iii) allow remote control and/or disconnection for credible risks to system security (for example to support frequency control)
>
>    iv) provide good-faith scheduling information to market and/or network operators (for example forecasted load/generation/storage profiles)

These responsibilities are outlined in R1.2.2, R2.2.2, and R3.2.2 for loads, generation and storage, and are included only where resources seek remuneration from energy market participation. That is, where resources do not seek remuneration from energy market participation, they should be subjected to only minimal obligations of technological capability, or connectivity. There is recognition that some loads may be required to be active based on threshold kW or kWh consumption values over a nominated period of time, for example an electric vehicle charging point, as a result of potential impact to grid operation. For customers who would wish to be able to use large electrical resources (e.g., rapid electric vehicle-chargers) or export electricity (e.g., through large PV-battery systems) without restriction, there is the possibility that customers or networks or the state would wish to invest in further network augmentation to allow this (for example by customers paying for three-phase power connections, or distribution businesses investing in network upgrades to support additional load/export capability). Public expectations of reasonable network capability will evolve over time and will remain an ongoing question for government consideration, but it is difficult to see how a blanket approach to allow unmitigated expectation of load/export capability would be either efficient or equitable. The ongoing development and uptake of algorithms and artificial intelligence to charge/discharge/export DER will likely further make augmentation costs prohibitively high in a DER-dominated future without the responsibilities for active participation outlined in R1.2.2, R2.2.2, and R3.2.2; uncoordinated algorithm-controlled charging of gigawatts of electric vehicle and battery charging will constitute a serious system security risk.

To note, these rights afford system and network operators the capability to set additional technical or participatory obligations on DER when active, for example through dynamic network agreements and dynamic operating envelopes. This is an intention of the Bill; a key value in framing active rights and responsibilities is that these may be implemented via a dynamic network agreement, without detailed consideration of operating envelope formulation or control hierarchies.

To note also, other than the above recognition of threshold values for load, the Bill does not distinguish for the size or rating of DER resources. This is an intentional omission to support self-consumption of self-generated and/or stored electricity with minimal interference or obligation. Current connection requirements in Australia often stipulate 5kW of export per phase as a reasonable size limit of connection, and whilst this may be both common and reasonable at the present time, it unclear what will be considered reasonable in the future. But even in this future, the guiding principles of a right to self-consumption with minimal interference should be preserved.

DER capability will likely span many orders of magnitude as commercial businesses follow residential customers in embracing DER and as more and more resources become capable of active participation. DER already are the largest generators by capacity in a number of jurisdictions, aggregated DER will likely be largest single 'units' in the near future. A further guiding principle of this Bill is that resources be treated symmetrically with large-scale resources where possible and efficient. That is, for resources that are actively participating in energy markets, exported energy and the provision of services should be remunerated proportionally to size, and with exposure to wholesale prices. This implies a principle to move away from guaranteed feed-in tariffs for exported energy and closer to export exposed to wholesale or market-contracted prices. The rapid reduction of state-legislated feed-in tariffs suggests this will rapidly become the case in the near future. This right to symmetric treatment is outlined in R1.2.4 and R2.2.3.

Data rights should be upheld for energy too and are explicitly recognised in the Consumer Data Right for Energy reforms currently underway in Australia. This program largely considers with third-party access to customer data and rights to support customer-switching and competition, but broader rights to privacy and data-access should additionally apply, and again with varying level of access that is proportional to active participation. These rights and responsibilities are outlined in R4.1.1 and R4.2.1.

Finally, the principle of default settings is conveyed in the Bill, recognising that passive options should be set as the default for DER (R5.1.1) and that informed customer choice is required for DER to participate actively in energy markets (R5.1.2). Additionally, that active options should only be enabled where there is net benefit to the customer (R5.2.1), providing explicit customer protection and confidence in behaviour by their retailer, aggregator and/or financially responsible market participant. The Bill outlines the right to revert to passive settings with associated obligations, to again support the ability for self-consumption of self-generated electricity, and use of passive reasonable loads with minimal interference. Passive settings may also be the default setting for DER in the event of loss of communication of DER, failure of compliance, or breach of cyber security (R5.2.3) – all growing concerns for system operators.

Overall, and perhaps most importantly, the Bill is drafted as simply as possible and in plain English to support customer interaction and endorsement.

**5.2 Implementation Pathway**

<u>Implementation instruments</u>

There are very many and very diverse regulatory instruments for energy participation in most countries ranging from product technical standards to network connection agreements, market ancillary service specifications, customer frameworks, national electricity laws, rules, guidelines and procedures. This draft Bill does not seek to be the implementation instrument for each of these rights but rather a document to guide individual implementation decisions and reforms within acceptable and endorsed parameters of customer interaction.

A number of existing instruments currently breach the proposed rights. For example, voltage response modes in the Australian Standard for grid-connected inverters (AS 4777.2) stipulate Volt-Watt curtailment at times of high network voltages that reduces not only export of residential solar PV generated electricity but also the possibility of self-consumption of this electricity.

This may be rectified and allowed to be placed in accordance with rights to self-consumption of self-generated electricity (R2.1.2) of the proposed Bill by inclusion of a 'Self-Consumption' clause in the standard, first proposed by P. Kilby of Energy Queensland (Figure 1).

There is currently limited representation of customers on the Australian Standards EL-042-03 Committee for the regular review of this standard, and as a result, minimal advocacy for the inclusion of this clause within the standard, which is resulting in a loss in value to customers of $1.2-4.5m per year in South Australia alone (Stringer et al. 2021). There is possibility of updated governance of standards for DER in Australia in the near future, but even these arrangements and deliberations are likely to be prohibitively detailed for broad customer interaction and endorsement.

With a nationally adopted DER Bill of Rights and Responsibilities, the EL-042-03 Committee may proceed with its current mandate to review and update the AS 4777.2 standard but with the requirement to ensure any changes to the standard do not breach defined rights. Similar consideration to support self-consumption of self-generated electricity may be applied to the use of Demand Response modes outlined in AS 4755 and Minimum System Load procedures by AEMO.

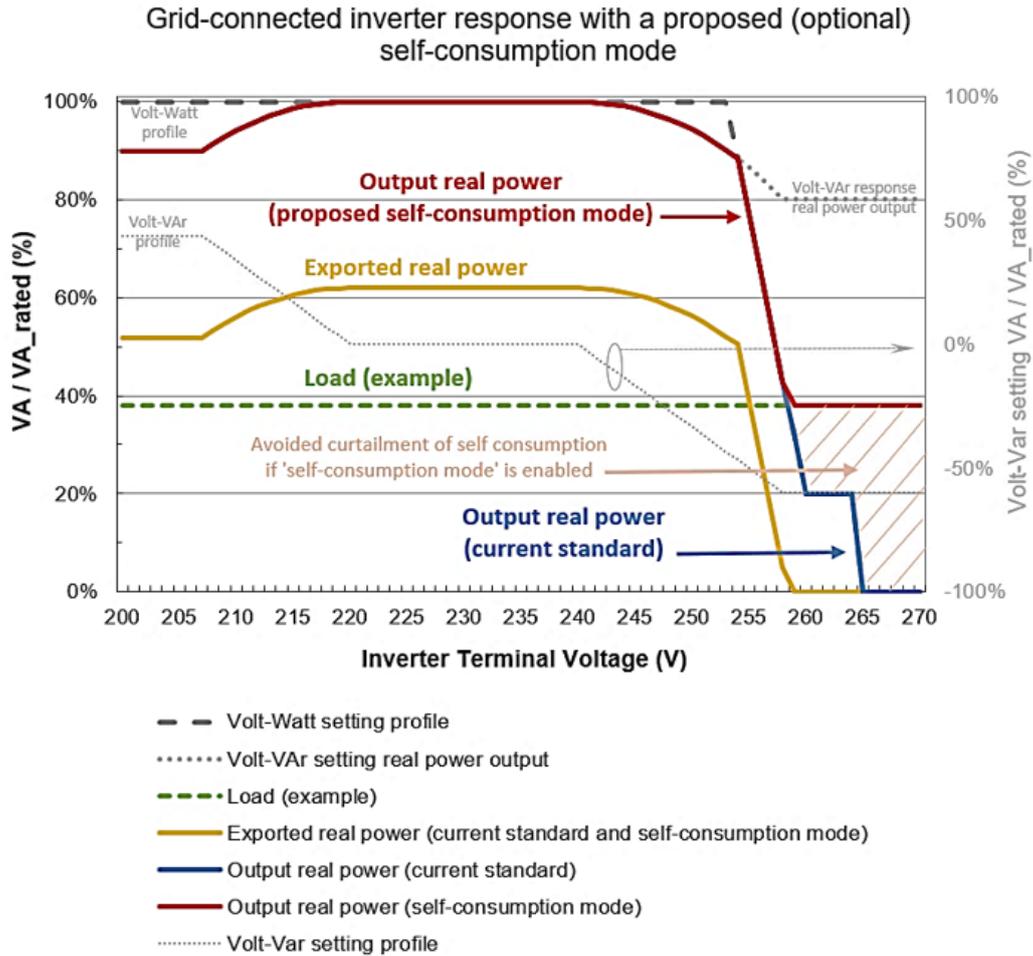

*Figure 1. A proposed self-consumption mode for grid-connected inverters to include in the Australian Standard AS 4777.2, after P. Kilby (pers. comm).*

Additional avenues for implementation of the proposed Bill include:

- Dynamic Network Agreements for control and communication of dynamic operating envelopes by distribution network service providers.
- Communication Protocols between market operators, distribution system operators, aggregators/traders, and individual devices – e.g., IEEE 2030.5 CSIP-AUS
- Consumer Data Right – Energy
- Ongoing reforms of DER integration led by the Australian Energy Security Board, including:
    - Scheduled Lite for additional visibility of non-scheduled distributed resources
    - Flexible Trader Arrangements to support communication, control, aggregation and accounting of multiple active and passive DER devices behind a single meter.

**5.3 Enshrinement pathway**

A national recognised DER Bill of Rights and Responsibilities would support swifter reforms of DER integration by giving confidence to consumers, their representatives, and the energy sector more broadly that reforms are occurring within agreed guiderails of customer-grid interaction. The most robust recognition of rights and responsibilities would be through legislative instruments, perhaps via state legislation or adoption through the National Electricity Rules, which would support legal recourse if any rights or responsibilities were breached. But there are non-legislative pathways that may facilitate steps towards endorsement and adoption in the interim.

The first necessary step is consideration and amendment of this Bill by consumer representatives. In Australia, leading customer advocates include the Public Interest Advocacy Centre, Australian Council of Social Services, St Vincent de Paul Society Victoria and the Energy Users Association of Australia. Additional to this are solar PV and DER advocates including Solar Citizens, Environment Victoria, Renew, Total Environment Centre, and industry peak bodies including the Smart Energy Council, the Clean Energy Council.

Following discussion, amendment and endorsement from these bodies, DER reform would be supported by the Bill being considered by jurisdictions and policy makers, market bodies, network service providers, retailers, aggregators and industry representatives, potentially through a formal engagement program such as the ESB Consumer Insights Forum and DER Implementation Plan. Following iterative review, discussion and endorsement amongst stakeholders, the Bill may then be explicitly referred to in reform workplans and policy deliberations. The guiding principles should remain that the Bill be intelligible, in plain English, support customer rights with passive reasonable engagement with electricity, and with fair allocation of responsibilities and benefits when customers choose to actively participate in energy markets.

## 6. Conclusion and Policy Implications

The energy revolution is proceeding at pace. Distributed energy resources are already comprising the majority of energy generation at times in some jurisdictions, likely to apply to all jurisdictions in Australia in the near future. A broad range of regulatory reforms are underway in response to this shift including decisions about emergency backstop PV control measures, solar export charges, dynamic operating envelopes, dynamic network agreements, flexible trader arrangements, interoperability standards, retail authorisations, access and pricing arrangements, trials and more.

Each of these reforms state a desire to place the customer at the centre of reforms, but the breadth and technical detail of each is limiting progress and customer endorsement and there is insufficient expertise and bandwidth of consumer advocates to meaningfully engage across all programs.

Whilst there are customer-focused updates to electricity rules and reviews of the National Energy Customer Framework and development of various industry codes of conduct, there is absent a clear, intelligible communication of customer rights and responsibilities for participation in the DER dominated energy system of the future. Particularly absent is a delineation between passive and active participation in energy markets.

With the draft DER Bill of Rights and Responsibilities outlined in this document we propose a first attempt at clarifying the rights and responsibilities for DER participation with the electricity grid in for four areas: 1) Energy consumption 2) Energy generation and 3) Energy storage and 4) Energy data. Each area has rights and responsibilities outlined for two broad categories of interaction 'Passive' and 'Active', and are informed by guiding principles including:

- To support system security and reliability with high DER penetration.
- To preserve the precedent of fair access to energy use by reasonable passive loads.
- To allow the self-use of self-generated electricity with minimum interference or obligation (with curtailment of self-consumption not exceeding the value of unserved energy for passive loads).
- To outline that where customers wish to interact with energy markets for remuneration though import, export or the provision of system services, they may be subjected to additional obligations of technical standards for grid support, information provision, and remote control or disconnection.

- That resources be treated symmetrically with large-scale resources where possible and efficient, and that some resources may be required to connect actively based on threshold kW or kWh values.
- That customers have a right to privacy and access to fair share of value from their energy data.
- That passive options should be set as the default for DER, with informed customer choice required for DER to participate actively in energy markets, and that active options should only be enabled where there is net benefit to the customer.

These rights are not limited in scope to household-scale DER; distributed energy resources will eventually be deployed at various scales throughout the grid and this Bill is constructed without arbitrary distinction of resource size.

Whilst not prescribing implementation for each of the rights and responsibilities, we highlight how various elements may be implemented through existing standards and processes, and how some rights are already being breached.

Finally, we outline possible pathways to enshrinement of these rights in a legislative instrument. The policy implication of this work is that the simple espousal of rights and responsibilities in plain English will support customers and their advocates to trust regulatory reform processes which abide by the Bill and/or for which there is a legislated avenue of appeal, and in doing so support swift, broadly endorsed, vital energy sector reform by providing clarity on the guiderails of DER interaction. As a result of its very high solar PV penetration Australia is required to consider these issues in detail now; it is likely that other countries will need to do so in the near future.

**Disclaimer**

Stocks, M., Blakers, A., Baldwin, K., 'Australia is the runaway global leader in building renewable energy', The Conversation, https://theconversation.com/australia-is-the-runaway-global-leader-in-building-new-renewable-energy-123694 accessed 1 October 2021

Stringer, N. et al. (2021), Renewable Energy 173, pp. 972

# Appendix A

An Australian Bill of DER Rights and Responsibilities Version 1, *Lal and Brown, 2022*

## 1. Energy Consumption

Passive consumption

1.1.1   Customers have a right to be connected to the grid and consume energy at fair prices without undue restriction or charges, where the consumption is occurring through reasonable use of passive loads.
1.1.2   Electricity supply for passive loads should meet the Reliability Standard

*Passive loads are considered those that do not respond to market signals or external coordination. Examples include televisions, hair-dryers, and dialysis machines.*

Active consumption

1.2.1 Customers have a right to own active loads that participate in energy markets

1.2.2 Where active loads are connected to the grid, they may be subjected to obligations to:

i) meet additional technical standards for active grid support

ii) register resource details at time of connection

iii)  allow remote control and/or disconnection for credible risks to system security

iv) provide good-faith scheduling information to market and/or network operators.

1.2.3 Active loads should be treated and remunerated symmetrically with large-scale loads where possible and efficient.

1.2.4 Customers have a right to switch active loads back to passive settings.*

*\*Some resources may be required to be active (e.g. electric vehicle charging points), based on threshold kW or kWh consumption over certain timeframes.*

## 2. Energy generation

Customers have a right to install renewable energy generation resources on their premises, subject to relevant requirements for safety and quality.

Self-consumption

2.1.1 Customers have a right to consume energy generated on-site with minimal restriction.

2.1.2 Energy generation systems may be subjected to obligations to:

i) meet minimum technical standards for passive grid support

ii) register system details at time of connection

iii) provide visibility of system operation to market and/or network operators

iv)  allow remote control and/or disconnection for non-credible risks to system security

2.1.3 Curtailment of self-consumption to not exceed the unserved energy provision in the Reliability Standard.

Export

2.2.1 Customers have a right to own generation systems that participate in energy markets

2.2.2 Energy generation systems capable of participating in energy markets may be subjected to additional obligations to:

i) meet additional technical standards for active grid support

ii) provide details of system operation

iii) provide forecasts of system operation to market and/or network operators

iv) allow adjustment and curtailment of exported energy to support system and network performance

v) allow remote control and/or disconnection for credible risks to system security

2.2.3. Generating systems that participate in energy markets have a right to symmetric treatment and remuneration with large-scale generation where possible and efficient

2.2.4 Customers have a right to switch actively participating systems back to passive settings, with minimal restriction of operation.

**3. Energy storage**

3.1.1 Customers have a right to install energy storage resources on their premises, subject to relevant requirements for safety.

Storage of energy generated on-site

3.1.2 Customers have a right to store energy generated on-site for later use

3.1.3 Storage systems may be subjected to obligations to:

i) meet minimum technical standards for grid support

ii) register system details at time of connection

iii) provide visibility of system operation to market and/or network operators

Storage of energy from the grid

2.2.1 Customers have a right to own generation systems that participate in energy markets

3.2.1 Customers have a right to own storage systems that participate in energy markets

3.2.2 Where capable of storing energy from the grid, operating as active load or participating in energy markets, storage systems may be subjected to additional obligations to:

ii) provide visibility of system operation to market and/or network operators

iii) allow remote control and/or disconnection for non-credible risks to system security

iv) provide good-faith scheduling information to market and/or network operators

v) allow remote control and/or disconnection for credible risks to system security

## 4. Energy data

4.1.1 Customers have a right to privacy through their energy data

4.1.2 Customers have a right to access their energy data

4.1.3 Customers have a right to be informed about the uses of their energy data

4.1.4 Customers have a right to a fair share in the value of third-party licensing of their energy data

4.1.5 Customers have a right to revoke consent of the use of their energy data

4.2.1 Where customers are actively interacting with the grid, they may have obligations to provide energy data relating to resource and/or system details, current operation, and forecast operation.

## 5. Default settings

5.1.1 Passive options are set as default for DER

5.1.2 Informed customer choice is required for DER to participate actively in energy markets

5.2.1 Active options should only be enabled where there is net benefit to the customer.

5.2.2 Customers have a right to revert to passive settings with associated obligations

5.2.3 Passive settings may be reverted to in the event of:

    i) Loss of communication with DER

    ii) Failure of compliance by DER

    iii) Cybersecurity compromise

    —